# Magnetic phase diagram of $Fe_{1.1}Te_{1-x}Se_x$: A comparative study with the stoichiometric superconducting $FeTe_{1-x}Se_x$ system

P. L. Paulose,[*] C. S. Yadav, and K. M. Subhedar

*Department of Condensed Matter Physics and Material Science, Tata Institute of Fundamental Research, Mumbai 400005, India.*

We report a comparative study of the series $Fe_{1.1}Te_{1-x}Se_x$ and the stoichiometric $FeTe_{1-x}Se_x$ to bring out the difference in their magnetic, superconducting and electronic properties. The $Fe_{1.1}Te_{1-x}Se_x$ series is found to be magnetic and its microscopic properties are elucidated through Mössbauer spectroscopy. The magnetic phase diagram of $Fe_{1.1}Te_{1-x}Se_x$ is traced out and it shows the emergence of spin glass state when the antiferromagnetic state is destabilized by the Se substitution. The isomer shift and quadrupolar splitting obtained from the Mössbauer spectroscopy clearly brings out the electronic differences in these two series.

The discovery of high temperature superconductivity in cuprate superconductors twenty years ago led to a vigorous exploration of possible novel mechanisms of pairing by electrons. The recent discovery of iron based superconductors has given a new impetus to this activity [1-8], bringing the magnetic spin fluctuation mediated pairing to the fore. One common feature observed in the Fe pnictides or chalcogenides is the gradual emergence of the superconductivity when the antiferromagnetic (AFM) order associated with the Fe-X (X=As/Te) layers is weakened by doping. The experiments and theoretical models suggest that in this class of materials the magnetic moments may be soft and depend sensitively on various physical details, which contrasts with the strongly correlated local behavior found in cuprates [9,10]. $Fe_{1+y}Te$ has a tetragonal anti-PbO type structure and forms only in the nonstoichiometric form (y in the range 0.05 to 0.22) with the excess iron atoms Fe(2), occupying the octahedral positions randomly. The Fe(2) directly couples to the four nearest neighbor Fe(1) atoms in the Fe planes and could effectively introduce frustration in the underlying antiferromagnetic state as Te is gradually substituted by Se and could lead to a large critical region [11]. It has also been suggested that the excess Fe bearing a local moment in proximity to the Fe layers could offer an interesting opportunity for experimental investigation of the interplay between superconductivity and pair breaking magnetic scattering in the Fe superconductors [12]. While $Fe_{1+y}Te$ orders antiferromagnetically, FeSe could be prepared in the stoichiometric form and is superconducting below 8K [10]. Density functional theory suggests that FeSe could be an antiferromagnet at the borderline between itinerant and localized behavior and the in-plane Fe-Fe exchange coupling depends on Fe-Se distance [13]. These predictions motivated us to undertake a systematic investigation using the microscopic Mössbauer spectroscopic and bulk magnetic studies of the Fe excess compounds $Fe_{1.1}Te_{1-x}Se_x$ and the stoichiometric $FeTe_{1-x}Se_x$. Our results show that $Fe_{1.1}Te_{1-x}Se_x$ ($0.1 \leq x \leq 0.55$) system is magnetic with an indication of well developed local moment, while the stoichiometric series $FeTe_{1-x}Se_x$ display bulk superconductivity with no magnetic ordering.

$Fe_{1.1}Te_{1-x}Se_x$ (x = 0, 0.05, 0.1, 0.2, 0.3, 0.4, 0.5, 0.55 and 1) and $FeTe_{1-x}Se_x$ (x=0.1, 0.3, 0.4, 0.5) samples were prepared by mixing the stoichiometric quantities of constituent elements of purity better than 99.9% and sintering at 700 $^0$C. The former set of samples will be referred as $Fe_{1.1}$ series and the latter set as $Fe_1$ series. $^{57}$Fe Mössbauer spectra were recorded in transmission geometry using a conventional constant-acceleration spectrometer and a helium flow cryostat. Isomer shift values are quoted relative to α-Fe at 295 K.

The Rietveld analysis (FULLPROF program) of the x-ray diffraction data shows that the samples are single phase (space group: *P4/nmm*). For the Fe excess series, the Rietveld fit and EDX analysis yielded the

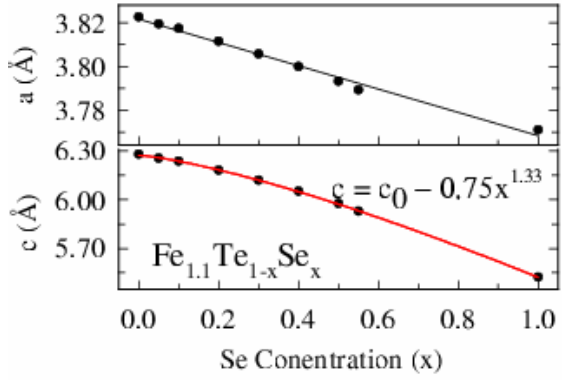

FIG. 1 (color online). The lattice parameters 'a' and 'c' as a function of Se concentration.

actual Fe composition as 1.10±0.02 while for the $Fe_1$ series, near stoichiometry was indicated within the detection accuracy of 0.02 by EDX. The lattice volume (V) and the lattice parameter 'a' vary almost linearly with Se concentration, while the parameter 'c' has a power law variation as $x^{1.33}$ (Fig.1). The lattice parameters for the stoichiometric series are very close to those of Fe excess series. However, we find a noteworthy feature, namely for x=0.3 compounds in the two series, 'a' is unchanged while 'c' is shortened by 0.005Å for the stoichiometric compound.

Figure 2(a,b) shows the electrical resistivity, ρ(T) for all the samples in both the series. For $Fe_{1.1}Te$, the ρ(T) shows a drop at 65K corresponding to antiferromagnetic (AF) ordering. Substitution of 5 at.% of Se for Te in $Fe_{1.1}Te$, lowers this transition temperature and broadens it considerably. Further increase of Se (x≥0.1) in both $Fe_{1.1}Te_{1-x}Se_x$ and $FeTe_{1-x}Se_x$ series, induces onset of superconductivity in the temperature range 12 K to 15 K. A maximum of $T_c$ occurs around 15 K for x=0.4 in both the series. For x≥0.3, the superconducting transition is sharper in the stoichiometric series compared to the Fe excess series. The resistivity above $T_c/T_N$ shows negative temperature coefficient with a logarithmic variation for the $Fe_{1.1}Te_{1-x}Se_x$ series. Our data shows a clear increase in the slope of the logarithmic ρ(T) behavior as the Se concentration increases. However, the stochiometric series show metallic behavior between $T_c$ and about 130 K for x≥0.3, with a logarithmic dependence [Fig 2(b) and inset]. Logarithmic dependence of ρ(T) has been

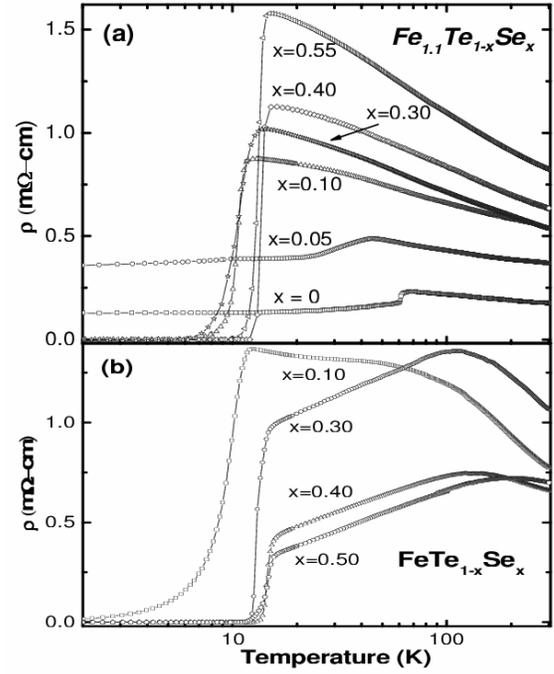

FIG. 2. Electrical resistivity curves:
(a) $Fe_{1.1}Te_{1-x}Se_x$ series (b) $FeTe_{1-x}Se_x$

reported for single crystalline $Fe_{1.04}Te_{0.6}Se_{0.4}$ and our polycrystalline data has similar absolute value of ρ at all temperatures [14]. For x=0.1, the ρ(T) curve displays no metallic behavior. As the Se doping increases there is an emergence of logarithmic metallic behavior which is reminiscent of J- positive Kondo effect [15]. The onset of this behavior progressively increases towards higher temperatures reaching to 170 K for x=0.5 [Fig. 2(b)].

We have performed dc magnetization measurements in an applied field of 0.5 T to investigate bulk magnetic order in these systems. For the $Fe_1$ series, the dc magnetization is nearly constant above $T_c$ (not shown here) implying that Fe is either nonmagnetic or weakly magnetic. Figure 3 shows the dc magnetization for the $Fe_{1.1}$ series. For $Fe_{1.1}Te$, a sharp drop in magnetization occurs below 65K corresponding to the AFM transition. Substitution of 5 at.% Se lowers this transition to 49K along with a considerable broadening. It is noteworthy that the electric transport faithfully captures the magnetic transition and its width for x=0 and 0.05 compositions [shown in Fig. 2(a)]. For x=0.1, the dc magnetization displays a clear cusp at 36K (Fig. 3). With further substitution

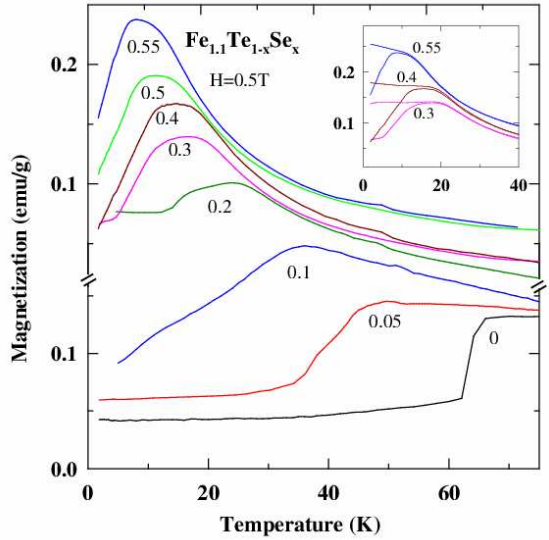

FIG. 3 (color online). M (T) at H=0.5T in the Zero field cooled (ZFC) state, the value of 'x' is shown near each curve. The inset shows FC and ZFC curves for x=0.3, 0.4 and 0.55.

of Se, the magnetization shows broad maximum that progressively shifts to lower temperatures. This type of magnetization behavior is typical of spin glass (SG) like systems. This SG nature is confirmed from sharp cusp in the $\chi_{ac}$ data, identified as the spin glass temperature, $T_g$ [inset of Fig. 4(a)]. The $\chi_{ac}$ peak for x=0.55 shifts to higher temperatures at higher frequencies proving the SG nature beyond doubt. Magnetic field cooling effects are also observed below the spin glass transition temperature $T_g$ (inset Fig. 3). $Fe_{1.1}Te$ has an ordered moment of about $1.8\mu_B$ and a paramagnetic effective moment ($\mu_{eff}$) of about $2.4\mu_B$ which indicates a localized spin state close to S=1 at every site [16]. From the Curie-Weiss fit to the M(T) data, we find that substitution of 20 at.% of Se at Te site lowers the $\mu_{eff}$ to $1.2\mu_B$ and further down to $0.8\mu_B$ with 55 at.% Se doping. This reduction of Fe moment is further supported by the microscopic study of Fe hyperfine field (Fig. 9).

We have carried out ac susceptibility studies to probe the details of magnetic and superconducting transitions (Fig. 4). The $Fe_{1.1}Te_{1-x}Se_x$ series with x≥0.1 show diamagnetism below the respective $T_c$. In the case of the $FeTe_{1-x}Se_x$ samples with x=0.3, 0.4 and 0.5 the diamagnetic screening corresponds

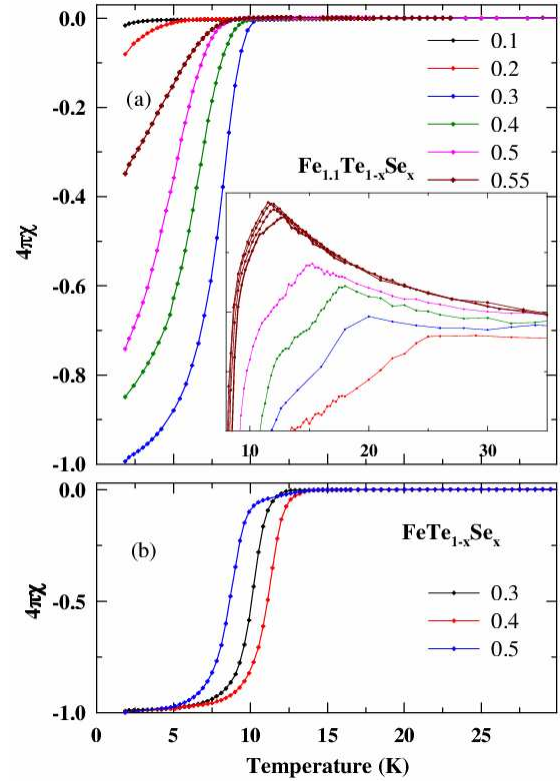

FIG. 4 (color online). (a) $\chi_{ac}$ curves at 10Hz for $Fe_{1.1}Te_{1-x}Se_x$ samples. The inset is a magnified part showing the SG transition; for x=0.55, $\chi_{ac}$ curves at 0.2Hz, 1Hz, 10Hz and 111Hz are shown (b) $\chi_{ac}$ at 10 Hz for $FeTe_{1-x}Se_x$.

to full value of $4\pi\chi$. Furthermore, the transition widths are narrower compared to the $Fe_{1.1}$ series [Fig. 4(b)].

The magnetic hysteresis studies show that the upper critical field ($H_{c2}$) is much greater than 12T for the superconducting samples in both the series. However, the critical current computed from the hysteresis loop at 1.8K is much higher for the stoichiometric series compared to the $Fe_{1.1}$ series. The heat capacity (C) measurements did not show any anomaly near $T_c$ for the samples in the $Fe_{1.1}$ series, whereas the C curves showed a very clear lambda anomaly at $T_c$ for the stoichiometric series for x=0.3, 0.4 and 0.5 and it persists up to a magnetic field of 14 T. Typical result is shown for $FeTe_{0.7}Se_{0.3}$ in figure 5. The inset in figure 5 shows the exponential drop indicating a clear gap in the superconducting state in these samples.

$^{57}Fe$ Mössbauer spectroscopy studies were carried out to probe the local magnetic state of Fe in these systems. The observed spectra

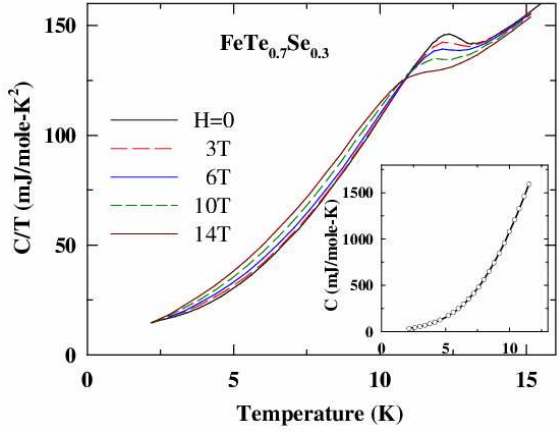

FIG. 5 (color online). Heat capacity at different applied fields. Inset is a fit for C at H=0.

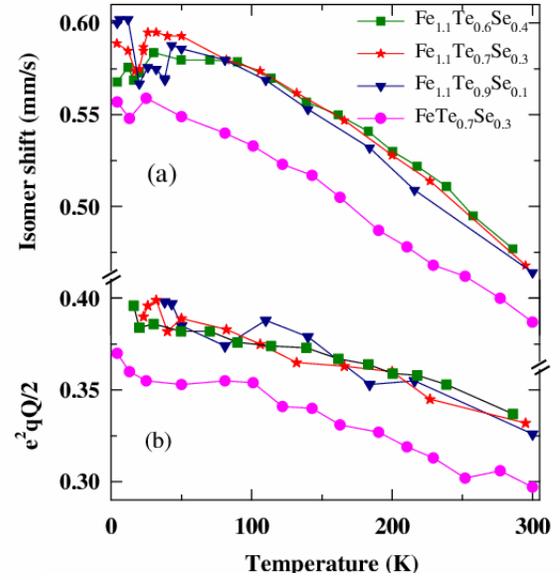

FIG. 6 (color online). Temperature variation of (a)Isomer shift, (b) Quadrupole splitting.

were least square fitted to yield the hyperfine parameters, namely isomer shift (IS), quadrupolar shift (QS) and the magnetic hyperfine field ($H_{hf}$). The spectra at 300K can be described by a single quadrupole doublet for all samples in the two series. This confirms that there is no magnetic ordering of any kind at room temperature and also rules out the presence of any magnetic impurity phase. However, it is to be noted that the line-width is more than the expected natural Fe line-width by about 20% and could be caused either by different Fe sites or inhomogeneities of various kinds. The Mössbauer isomer shift (IS) is found to increase as the temperature is decreased (Fig. 6). The total change between 300K and 5K is about 0.12mm/s. This variation can be mainly attributed to the relativistic temperature-dependent contribution to the isomer shift caused by the second order Doppler shift of the emitted γ-rays. A significant find is that the IS curves nearly overlap for the samples in the $Fe_{1.1}$ series, while it is shifted to lower values by about 0.04mm/s for the $Fe_1$ series (only x=0.3 curve is shown). This shows a clear electronic difference between these two series. A diminished screening of 4s electrons caused by a decrease in '*d*' electrons at Fe can explain the decrease in isomer shift. The quadrupole splitting (QS) shows an increase with decreasing temperature for all the samples in both the series and can be attributed to local distortion from the tetragonal symmetry at the Fe site. Surprisingly the QS curves overlap for the $Fe_{1.1}$ series though the Se concentration changes. However the QS curve for $Fe_1$ series is lowered by about 0.03mm/s. An increase in QS compared to the stoichiometric series may imply an enhancement of lattice distortion by the excess Fe.

For $FeTe_{1-x}Se_x$ series, the single quadrupole paramagnetic doublet does not acquire much additional broadening down to 4.2 K (Fig. 7). This observation coupled with the magnetization data shows that Fe(1) is nonmagnetic in the $Fe_1$ series. However for the $Fe_{1.1}$ series, additional broadening of the line is observed on lowering the temperature and can be attributed to magnetic hyperfine field. We find that the temperature, at which the broadening sets in, coincides with the respective $T_g$ observed from the $\chi_{ac}$ studies. This proves the magnetic origin of the broadening. Further, the broadening at 4.2K is found to decrease with increase in Se concentration. This trend correlates well with the systematic decrease in $T_g$ and the magnetic moment on Fe as indicated by $\mu_{eff}$. These observations conclusively show that the spin glass ordering is global in nature and not from any possible minor impurity phases. The spectra at 4.2K are very broad and not well resolved (Fig. 7). They were analyzed using the Window method where the magnetic hyperfine field distribution is assumed to be a Fourier series with a finite number of terms [17]. The probability distribution of hyperfine field obtained is shown in Fig. 8. The distribution is very broad, revealing the

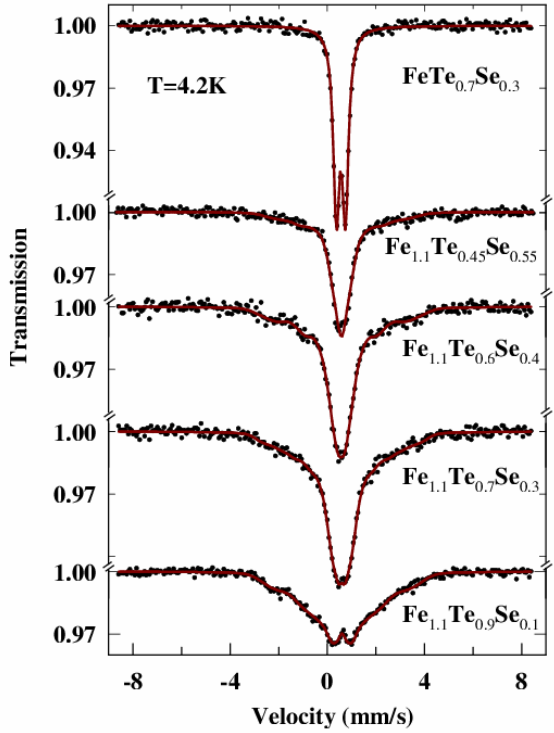

FIG. 7 (color online). Mössbauer spectra at 4.2 K for selected samples.

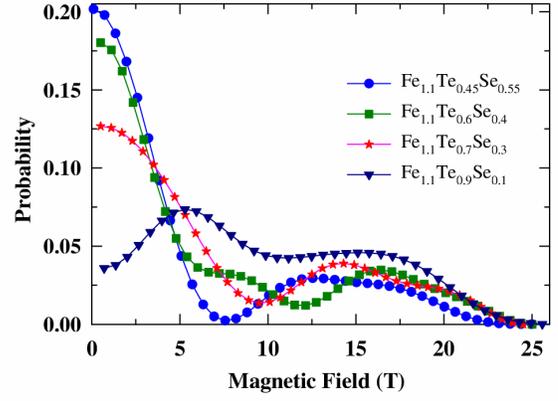

FIG. 8 (color online). Hyperfine field distribution from the Mössbauer spectra $Fe_{1.1}Te_{1-x}Se_x$ (x=0.1, 0.3, 0.4, 0.55).

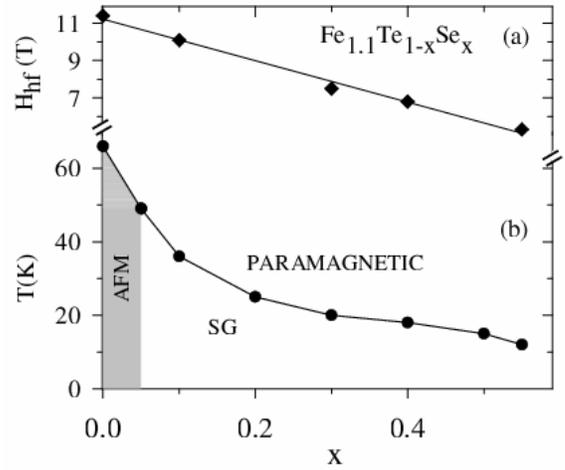

FIG. 9. (a) Variation of average hyperfine field with Se concentration (b) Magnetic phase diagram

sensitive nature of the Fe hyperfine field to its local environment in this system. The distribution is nearly bimodal with a high field component centered around 15T, possibly representing Se deficient neighborhood. The low field component is strongly enhanced by the Se substitution and this could be a contributing factor for the emergence of the competing magnetic interactions leading to the spin-glass state. The average hyperfine field ($H_{av}$) is found to drop linearly with the Se substitution (Fig. 9) indicating that Se substitution gradually decreases the Fe moment, in agreement with the bulk magnetization data. The fit procedure treats the central quadrupole doublet corresponding to essentially non-magnetic iron as if it were a magnetic hyperfine pattern with $H_{av} < 4$ Tesla. The magnetic fraction of Fe atoms are estimated from the hyperfine distribution excluding the region up to 4 Tesla. The fractional values obtained are 0.75, 0.5, 0.43 and 0.32 for x=0.1, 0.3, 0.4 and 0.55 respectively and it clearly exceeds Fe(2) fraction which is only 10%. This indicates that Fe(2) in the octahedral sites has strong direct, magnetic coupling with the Fe(1) in the planes. The progressive reduction in the magnetic fraction of the Fe atoms can be linked to the reduction of $\mu_{eff}$ with the substitution of Se for Te.

Based on these results we arrive at the magnetic phase diagram of the $Fe_{1.1}Te_{1-x}Se_x$ series as shown in figure 9. It is evident that the AF state is destroyed beyond 5at.% Se at Te site and the system transforms into a spin glass state. This phase diagram can be understood if we assume a model where Fe(1) planes are made weakly magnetic as a result of Se substitution while Fe(2) sites remain strongly magnetic and introduces random competing exchange interactions. The excess Fe, i.e. Fe(2) displays local magnetism and interacts with Fe(1), thereby strongly influences the superconductivity caused by the Fe(1) layers.

In conclusion, we have presented systematic studies of stoichiometric $FeTe_{1-x}Se_x$ and $Fe_{1.1}Te_{1-x}Se_x$. It brings out the role of excess Fe in the $Fe_{1+y}(TeSe)$ system which was so far speculative. Our data gives strong evidence that Fe(2) interacts magnetically with Fe(1) responsible for superconductivity. The superconducting transitions observed from the electrical resistivity and magnetic susceptibility in the $Fe_{1.1}Te_{1-x}Se_x$ series are not supported by the heat capacity data and may be indicative of weak filamentary superconductivity and one has to be cautious about using such samples for superconductivity studies. However there may be a weak case for the coexistence of spin glass order and superconductivity in the $Fe_{1.1}Te_{1-x}Se_x$ series for x>0.1, but needs a more detailed study. The IS and QS behavior show that there is a clear electronic difference in the two series. The stoichiometric series is bulk superconductors for x=0.3,0.4 and 0.5 and the Fe Mössbauer line indicates no magnetic features down to 4.2K.

*E-mail: paulose@tifr.res.in

[1] KAMIHARA Y. *et. al.*, *J. Am. Chem. Soc.* **128** (2006) 10012.
[2] KAMIHARA Y. *et. al.*, *J. Am. Chem. Soc.* **130** (2008) 3296.
[3] CHEN X.H *et. al.*, *Nature* **453** (2008) 761.
[4] CHEN G.F. *et. al.*, *Phys. Rev. Lett.* **100** (2008) 247002.
[5] REN ZHI-AN *et. al.*, *Europhys. Lett.* **83** (2008) 17002.
[6] HSU F. C. *et. al.*, *Proc. Natl. Acad Sci. U.S.A.* **105** (2008) 14262.
[7] FANG M.H. *et. al.*, *Phys. Rev. B* **78** (2008) 224503.
[8] SALES B. C. *et. al.*, *Phys. Rev. B* **79** (2009) 94521.
[9] MAZIN I. and JOHANNES M.D., *Nature Physics* **5** (2009) 141.
[10] McQUEEN T. M. *et. al.*, *Phys. Rev. B* **79** (2009) 014522.
[11] FANG CHEN *et. al.*, *Europhys. Lett.* **86** (2009) 67005
[12] ZHANG L. *et. al.*, *Phys. Rev. B* **79** (2009) 012506.
[13] PULIKKOTIL J. J. *et. al.*, *ArXiv*:0809.0283v1.
[14] LIU T.J.*et. al.*, *Phys. Rev. B* **80**(2009)174509
[15] KONDO J., *Prog. Theor. Phys.* **32** (1964) 37.
[16] TSUBOKAWA I. and CHIBA S., *J. Phys. Soc. Japan* **14** (1959) 1120.
[17] WINDOW B., *J. Phys. E* **4** (1971) 401.